\documentclass[prd,aps,floats]{revtex4}

\usepackage{latexsym}
\usepackage{amssymb}
\usepackage{amsmath}

\usepackage{epsfig}

\newcommand{\etal}{{\it et al.}}
\newcommand{\apjs}{Astrophys. J. Supp.}
\newcommand{\apjl}{Astrophys. J. Lett.}
\newcommand{\mnras}{Mon.Not.Roy.Astron.Soc.}

\newcommand{\connE}{\ensuremath{\tilde{\nabla}}}
\newcommand{\connM}{\ensuremath{\nabla}}
\newcommand{\metE}{\ensuremath{\tilde{g}}}
\newcommand{\metM}{\ensuremath{g}}

\newcommand{\volE}{\ensuremath{\sqrt{-\metE}}}
\newcommand{\volM}{\ensuremath{\sqrt{-\metM}}}
\newcommand{\RiemE}{\ensuremath{\tilde{R}}}

\newcommand{\EinE}{\ensuremath{\tilde{G}}}

\newcommand{\grad}{\ensuremath{\vec{\nabla}}}

\newcommand{\Vp}{\ensuremath{\frac{dV}}{d\mu}}
\newcommand{\Vpp}{\ensuremath{\frac{d^2V}}{d\mu^2}}

\newcommand{\XiE}{\ensuremath{\tilde{\Xi}}}
\newcommand{\chiE}{\ensuremath{\tilde{\chi}}}
\newcommand{\zetaE}{\ensuremath{\tilde{\zeta}}}
\newcommand{\nuE}{\ensuremath{\tilde{\nu}}}

\newcommand{\phib}{\ensuremath{\bar{\phi}}}
\newcommand{\mub}{\ensuremath{\bar{\mu}}}
\newcommand{\rhob}{\ensuremath{\bar{\rho}}}
\newcommand{\Pb}{\ensuremath{\bar{P}}}

\newcommand{\curv}[1]{\ensuremath{\frac{#1K}{r_c^2}}}

\begin{document}

\title{Generalizing TeVeS Cosmology}

\author{Constantinos Skordis}
\affiliation{ Perimeter Institute, Waterloo, Ontario N2L 2Y5, Canada.} 
\email{cskordis@perimeterinstitute.ca}

\date{\today}

\renewcommand{\thefootnote}{\arabic{footnote}} \setcounter{footnote}{0}

\begin{abstract}
I consider an extented version of Bekenstein's Tensor-Vector-Scalar theory where the action
of the vector field is of a general Einstein-Ether form. This work presents the cosmological equations of this
theory, both at the background and perturbed level, for scalar, vector and tensor perturbation modes.
By solving the background equations in the radiation era analytically, to an excellent approximation, 
I construct the primordial adiabatic perturbation for a general family of scalar field kinetic functions.
\end{abstract}

\maketitle
\section{Introduction}
In the last two decades, cosmology has undergone a "precision" revolution, with a large influx of data such as observations of the 
Cosmic Microwave Background~\cite{wmap}, large scale structure~\cite{LSS}, and supernovae observations~\cite{super}.
There is now a consensus cosmological standard model, named $\Lambda$CDM, based on General Relativity as the theory of gravity, which requires only about $4\%$ of
the energy budget of the universe to be in known baryonic form, while the rest is divided into two apparently distinct, dark
components: \emph{cold dark matter} and \emph{dark energy}. It is unfortunate however, that apart from their
phenomenology as fluids, we know nothing of their actual nature at the present time. 

Cold dark matter, typically composed of  very massive slowly moving
and weakly interacting particles,is required on cosmological scales mainly to source large scale structure. To dramatise its importance, assuming general relativity,
if dark matter was absent, structure as we know it would not have even formed yet.
A plethora of such particles generally arises in particle physics models beyond the standard model quite
naturally with the right cross-sections to create the right abundance (see ~\cite{particle_DM1,particle_DM2} for a reviews). Yet, 
while its phenomenology as a dust fluid has been shown to agree with observations to a very good degree,
 the actual nature of cold dark matter is left to speculation as no cold dark matter particle has been observed so far. 
Moreover there are still some mishaps within the $\Lambda$CDM paradigm, for example 
the problem of voids~\cite{void} and the recent observations of the Abel 520 cluster~\cite{abel}.

Given that the law of gravity plays a key role, to all observations from which dark matter and dark energy are inferred, it is
conceivable that General Relativity breaks down at small enough gradients and curvatures, and an alternative theory of gravity might also provide 
an explanation to the dark sector. One such theory proposed some time ago by Bekenstein~\cite{teves}, building on key work by Sanders~\cite{sanders_stratified}. 
This theory was dubbed Tensor-Vector-Scalar (TeVeS) because it relies on a bimetric transformation involving a scalar and a vector field.
It was designed to reduce to the Aquadratic Lagrangian non-relativistic
theory of Bekenstein and Milgrom~\cite{aqual}. Thus, it provides essentially the same phenomenology as Milgrom's Modified Newtonian Dynamics (MOND)~\cite{MOND}
 for galactic rotation curves, for which MOND has had tremendous success~\cite{MOND_galaxy}.

TeVeS theory has since been shown~\cite{SMFB} to be able to source structure in a similar way as dark matter. The vector field
in the theory plays a key role~\cite{DL}, as for a wide range of parameters it has a power-law growing mode which sources potential wells.
In contrast with dark matter, the vector field has shear which creates a mismatch between the two scalar gravitational potentials. This has been identified
as prospective discriminator between dark matter and theories like TeVeS~\cite{mismatch1,mismatch2,schmidt,zhang}. Other non-cosmological tests of TeVeS 
have also been studied, for example gravitational lensing~\cite{lensing1,lensing2}. Probing the difference between the arrival times of neutrinos and
gravitational waves from distant supernovae is also a possibility~\cite{kahya}.

Having shown that one can cast TeVeS in a single metric form, with the scalar field absorbed into the vector field~\cite{teves_single},
Zlosnik, Ferreira and Starkman, have explored a sister theory, based solely on a unit-timelike vector field with a non-canonical kinetic term~\cite{ZFS1} which
is a non-canonical Einstein-ether theory~\cite{einstein_ether}. This theory has also
been shown to source structure in a similar way as TeVeS~\cite{ZFS2} (cosmology within the context of canonical Einstein-ether theory has been extensively studied 
in~\cite{carroll,lim,LMB}) while its predictions for corrections to Newtonian gravity in the solar system have also been studied~\cite{bonvin}.
 Further exploration of this theory into different directions has been considered by Halle and Zhao~\cite{zhao,halle_zhao}. 

In this work I initiate a study of a version of TeVeS which involves a generalization of the action of the vector field into the Einstein-ether form.
This is motivated in part from the instability present in spherically symmetric solutions to the original TeVeS theory~\cite{seifert}, that possibly stems from
the fact that the kinetic term for the unit-timelike vector field was of Maxwellian form which can violate the dominant energy condition~\cite{skordis_dom}. Seifert has shown that
more general Einstein-ether actions can be stable depending on the parameters. It is therefore of importance to check whether forms of TeVeS exist which 
have stable spherically symmetric solution and still can form large scale structure. Possible directions for deriving TeVeS-type theories from
more fundamental theories are discussed in\cite{bekenstein_review,brunetton,mavromatos}.

In section-\ref{fund}, I lay down the action and derive the field equations.  The cosmological equations 
for a (possibly curved) Friedman-Lema\^{i}tre-Robertson-Walker (FLRW) metric are studied in section-\ref{FLRW}, where it is shown that they are identical with the TeVeS equations
upto a rescaling of Hubble's constant. The cosmological solutions are therefore the same as those in TeVeS. I conclude that section by deriving an
approximate (to an excellent degree) solution in the radiation era for a general family of scalar field functions.

To study large scale structure we need the cosmological perturbation equations about an FLRW universe. The gauge form-invariant perturbed equations are
derived using the techniques in~\cite{skordis_PT} and are given
in section-\ref{PT} for all types of perturbations, namely scalar vector and tensor perturbations.
Using the approximate background solution in the radiation era from section-\ref{FLRW},  I construct the primordial adiabatic perturbation. The construction
of the most general, regular primordial perturbation is quite involved and is given elsewhere~\cite{iso}. Throughout the paper I use the conventions of Wald~\cite{wald}.

\section{Fundamentals : action and field equations}
\label{fund}

\subsection{Preliminaries}
TeVeS theory and the generalization herein, is a bimetric theory where gravity is mediated by a tensor field $\metE_{ab}$ \label{def_metE} with 
associated metric-compatible connection $\connE_a$ and well
defined inverse $\metE^{ab}$ such that $\metE^{ac}\metE_{cb} = \delta^a_{\;\;b}$,
 a timelike (dual) vector field $A_a$ \label{def_A} such that  $\metE^{ab}A_a A_b = -1$,
and a scalar field $\phi$\label{def_phi}. Matter is required to obey the
weak equivalence principle, which means that there is a metric $\metM_{ab}$  \label{def_metM}  with
associated metric-compatible connection $\connM_a$, universal
to all matter fields, such that test particles follow its geodesics. The tensor field $\metE_{ab}$ will
be called the {\it Einstein-Hilbert frame metric} (see below) while $\metM_{ab}$ the
{\it matter frame metric}.  

The relation between the four above tensor fields (when the field equations are satisfied) is
\begin{equation}
   \metM_{ab} = e^{-2\phi}\metE_{ab} - 2\sinh(2\phi)A_a A_b
   \label{eq:metric_relation}
\end{equation}
with inverse
\begin{equation}
   \metM^{ab} = e^{2\phi}\metE^{ab} + 2\sinh(2\phi)A^a A^b
   \label{eq:inv_metric_relation}
\end{equation}
where  $A^a = \metE^{ab}A_b$.

\subsection{The action principle}
The theory is based on an action $S$, which splits as $S = S_g + S_A + S_{\phi}+S_m$, where 
$S_g$,$S_A$,$S_{\phi}$ and $S_m$ are the actions for  $\metE_{ab}$,
  vector field $A_a$, scalar field $\phi$ and matter respectively. 

The action for  $\metE_{ab}$, $A_a$ and $\phi$  is most easily written in the 
Einstein-Hilbert frame, and is such that $S_g$ is of Einstein-Hilbert form
\begin{equation}
   S_g = \frac{1}{16\pi G}\int d^4x \; \volE \; \RiemE,
\end{equation}
where $\metE$ and $\RiemE$ are \label{def_detE} \label{def_RE}  the determinant and scalar curvature of $\metE_{\mu\nu}$ respectively and
$G$ \label{def_Gbare} is the bare gravitational constant. Due to the complicated nature of the equations, the numerical value of $G$ will not
be the measured value of Newton's constant as measured on Earth. The precice relation between them depends on 
the spherically symmetric solution which apart from depending on the arbitrary function $V$ (see below), is not expected to be unique, just like the
case of standard TeVeS~\cite{teves,spher_sym_teves}. 

The action for the vector field $A_a$ is given by
\begin{equation}
	S_A = -\frac{1}{16\pi G}  \int d^4x \; \volE \; \left[ K^{abcd}\connE_a A_b \connE_c A_d   - 2\lambda (A_a A^a + 1)\right],
\end{equation}
where 
\begin{equation}
K^{abcd} = K_B\left(\metE^{ac}\metE^{bd} - \metE^{ad}\metE^{bc}\right) + K_+ \left( \metE^{ac}\metE^{bd} + \metE^{ad}\metE^{bc} \right) + K_0 \metE^{ab}\metE^{cd} + K_A \metE^{bd} A^a A^c
 \label{def_K}
\end{equation}
$\lambda$ \label{def_lambda} is a Lagrange multiplier ensuring the timelike constraint on $A_a$ and $K_B$, $K_+$, $K_0$ and $K_A$\label{def_KB} are  dimensionless constants.

The action for the scalar field $\phi$ is given by
\begin{equation}
    S_{\phi} = -\frac{1}{16\pi G} \int d^4x  \volE \left[ 
    \mu \left(\metE^{ab} - A^a A^b\right)\connE_a\phi \connE_b\phi +   V(\mu) \right]
\end{equation}
 where $\mu$ \label{def_mu} is a non-dynamical dimensionless scalar field,  and $V(\mu)$ is an arbitrary 
function which must be such that $\Vp\rightarrow \mu^2$ as $\mu\rightarrow 0$ in order to have exact MOND limit, while it
must diverge as $\mu \rightarrow \mu_0$ where $\mu_0$ is a constant, in order to have exact Newtonian limit~\cite{Bourliot}. One example is the form
considered in ~\cite{Bourliot} which is 
\begin{equation}
 \Vp = \frac{\mu_0^2}{16\pi \ell_B^2} \frac{\hat{\mu}^2}{\hat{\mu} - 1} (\hat{\mu} - \mu_a)^n
\label{bourliot_function}
\end{equation} 
where $\ell_B$ is a scale, $\mu_a$ is a constant,$n$ is an integer power and $\hat{\mu} = \frac{\mu}{\mu_0}$.
This general class of functions will also be used in this work.
 
The matter is coupled only to the matter frame metric $\metM_{ab}$ and thus its action is 
of the form 
$ S_m[\metM,\chi^A] = \int d^4x \; \volM \; L[ \metM ,\chi^A]$
for some generic collection of matter fields $\chi^A$. 

One can further generalize the vector field action by making the constants $K_B$, $K_+$, $K_0$ and $K_A$ functions of the scalar field $\phi$, as well as 
generalizing the scalar field action by inserting $\phi$-dependent functions as coefficients of the terms $\tilde{g}^{ab}$ and $A^a A^b$ in the
kinetic term   and  a potential for $\phi$ . I leave this for future investigations (if warranted).

\subsection{The field equations}
Variation wrt the  Lagrange multiplier $\lambda$ gives back the  timelike constraint on the vector field, $\metE^{ab}A_a A_b = -1$.
The  matter stress-energy tensor $T_{ab}$  is defined by
varying of the matter action w.r.t the matter frame metric as
  $\delta S_m = -\frac{1}{2} \int d^4x \volM \; T_{ab} \; \delta\metM^{ab}$.

Define tensors $ S^{efcd}_{\;\;\;\;\;\;ab}$ and $J^{abcde}$ as
\begin{eqnarray}
  S^{efcd}_{\;\;\;\;\;\;ab} =  \frac{\delta K^{efcd}}{\delta \metE^{ab}}
 &=&
  K_B \left[\delta^e_{\;\;(a}\delta^c_{\;\;b)} \metE^{fd} + \metE^{ec}\delta^f_{\;\;(a}\delta^d_{\;\;b)}  
  -\delta^e_{\;\;(a}\delta^d_{\;\;b)} \metE^{fc} - \delta^f_{\;\;(a}\delta^c_{\;\;b)} \metE^{ed} \right]
\nonumber \\
&&
 + K_+ \left[\delta^e_{\;\;(a}\delta^c_{\;\;b)} \metE^{fd} + \metE^{ec}\delta^f_{\;\;(a}\delta^d_{\;\;b)} 
  + \delta^e_{\;\;(a}\delta^d_{\;\;b)} \metE^{fc} + \delta^f_{\;\;(a}\delta^c_{\;\;b)} \metE^{ed} \right]
\nonumber \\
&&
  + K_0\left[ \delta^e_{\;\;(a}\delta^f_{\;\;b)}  \metE^{cd} + \delta^c_{\;\;(a}\delta^d_{\;\;b)} \metE^{ef} \right]
+ K_A  \left[\delta^f_{\;\;(a}\delta^d_{\;\;b)} A^e A^c + \metE^{fd} \delta^e_{\;\;(a} A_{b)} A^c + \metE^{fd} \delta^c_{\;\;(a} A_{b)} A^e \right]
\end{eqnarray}
and
\begin{equation}
J^{abcde} =  \frac{\delta K^{abcd}}{\delta A_e} =  K_A \metE^{bd} (\metE^{ae} A^c  + \metE^{ce} A^a )
\end{equation}
respectively.

Then the field equations for $\metE_{ab}$ are given by
\begin{eqnarray}
   \EinE_{ab} &=& 8\pi G\left[ T_{ab} + 2(1 - e^{-4\phi})A^c T_{c(a} A_{b)}\right]
              - \lambda A_a A_b
+ \mu \left[ \connE_a \phi \connE_b\phi  - 2 A^c\connE_c\phi A_{(a}\connE_{b)}\phi  \right]
        + \frac{1}{2}\left(\mu V' -  V\right) \metE_{ab}   
  \nonumber \\ 
&&        
+ \left[S^{efcd}_{\;\;\;\;\;\;\;ab} -\frac{1}{2}K^{efcd} g_{ab} \right]\connE_e A_f \;\connE_c A_d 
 - \connE_e \left[ \left( A_{(a} K_{\;\;b)}^{e\;\;\;cd} +  A_{(a} K_{b)}^{\;\;\;ecd} - A^e K_{(ab)}^{\;\;\;\;\;cd}\right)\connE_c A_d \right]
\end{eqnarray}
where $\EinE_{ab}$ is the Einstein tensor \label{def_G} of $\metE_{ab}$.

The field equations for the vector field $A_a$ are
\begin{equation}
	K^{abc}_{\;\;\;\;\;d} \connE_c \connE_a A_b 
= \left[\frac{1}{2}J^{abce}_{\;\;\;\;\;\;\;d} - J^{a\;\;ceb}_{\;\;d}\right]\connE_a A_b \connE_c A_e -\lambda A_d -
        \mu A^b\connE_b\phi \connE_d\phi  + 8\pi G  (1 - e^{-4\phi})A^b T_{ba} \label{eq:A_eq} 
\end{equation}

The field equation for the scalar field $\phi$ is 
\begin{eqnarray}
   \connE_a \left[   \mu \left(\metE^{ab} - A^a A^b\right)\connE_b\phi \right] &=& 8\pi G e^{-2\phi}\left[\metM^{ab} + 2e^{-2\phi} A^a A^b\right] T_{ab}
 \label{eq:Phi_eq}
\end{eqnarray}
where the non-dynamical field $\mu$ is found by inverting
\begin{equation}
      \left(\metE^{ab}- A^a A^b\right)\connE_a\phi \connE_b\phi  = -V' \label{eq:mu_con} 
\end{equation}
and therefore the arbitrary function $V$ and its derivatives
are nothing but functions of kinetic terms for $\phi$, contracted with $\metE^{ab}$ and $A^a$.

\section{FLRW cosmology}
\label{FLRW}

\subsection{Equations}
A most convenient coordinate system that is commonly used in cosmological perturbation theory
 is the conformal synchronous coordinate system with $t$ denoting conformal time and $x^{\hat{a}}$
 the spatial coordinates. 
This gives the matter frame metric with scale factor $a$ as  
\begin{equation}
  ds^2 = a^2\left[ -dt^2 + q_{ij} dx^{i} dx^{j}\right]
\end{equation}
where $q_{ab}$ is the metric of a space of constant curvature $\curv{}$, with radius of curvature $r_c$ and where
$K=0$ for a flat , $K=1$ for positively curved  and $K=-1$ for negatively curved space. The scale factor of the
Einstein-frame metric is $b = a e^{\phi}$.
The vanishing of the Lie derivative with respect to all the Killing vectors of the 
background spacetime gives $\phi = \bar{\phi}(t)$ only, while the vector field is pure gauge for this background.

The scalar field is governed by the TeVeS constraint which in this coordinate system reads,
\begin{equation}
   \dot{\bar{\phi}}^2 =  \frac{1}{2}a^2 e^{-2\phi} \Vp
\end{equation}
which must be inverted to get $\bar{\mu}(a,\bar{\phi},\dot{\bar{\phi}})$, and
the second order equation
\begin{equation}
\ddot{\bar{\phi}} =  \dot{\bar{\phi}} \left(\frac{\dot{a}}{a} - \dot{\bar{\phi}}\right) 
-  \frac{1}{U}\left[ 3 \bar{\mu}\frac{\dot{b}}{b}\dot{\bar{\phi}}
   + 4\pi G a^2 e^{-4\bar{\phi}} (\bar{\rho} + 3 \bar{P})\right], 
\end{equation}
where $U = \mu + 2 \Vp / \Vpp$.
Both of the above are unchanged from TeVeS, because they are not affected by the vector field action.
  
Defining the constant $K_F = 1 + K_0 + \frac{3}{2} K_+$, the Friedmann equation gives
\begin{equation}
 3 K_F \frac{\dot{b}^2}{b^2}
 = a^2e^{-4\phi}\left[\frac{1}{2} e^{2\phi} (\mu  \Vp + V)  +  8\pi G \bar{\rho} -  \frac{3K}{r_c^2a^2} \right]
\end{equation}
while the Raychandhuri equation  is
\begin{equation}
   K_F \left[- 2 \frac{\ddot{b}}{b} + \frac{\dot{b}^2}{b^2} -  4 \frac{\dot{b}}{b}\dot{\bar{\phi}}\right]   = 
 a^2 e^{-4\phi}\left[\frac{1}{2}e^{2\bar{\phi}}\left(\mu \Vp - V\right) +  8\pi G\bar{P} + \frac{K}{r_c^2a^2}\right] 
\end{equation}
where $\bar{\rho}$ and $\bar{P}$ are the energy density and pressure of a matter fluid, and evolve as
\begin{equation}
   \dot{\bar{\rho}} + 3\frac{\dot{a}}{a}(1 + w) \bar{\rho} = 0.
\end{equation}
with $w = \bar{P}/\bar{\rho}$.

\subsection{Solution in the radiation era}
\label{back_rad}
For the form of the function above (\ref{bourliot_function}),  it has been shown~\cite{SMFB,DL,Bourliot} that the scalar field tracks the
dominant fluid such that the energy density of the fluid relative to the energy density in the scalar field is constant. This tracker behaviour is found
in many scalar field dark energy (quintessence) models~\cite{quinte}. However it has also been shown~\cite{DL} for 
the special case of the Bekenstein toy model ($n=2$ and $\mu_a = 2$ in the function above), that in the radiation era for realistic baryon and radiation 
densities~\footnote{By radiation, I mean the total contribution from all relativistic species, like photons, neutrinos etc.}
the radiation tracker is almost never reached until just before the transition to the matter era. 
Instead the solution is such that $\dot{\phi}$ evolves as a powerlaw of the scale factor.
I show here that this is also true for the generalized function  above.

In the deep radiation era, for the function (\ref{bourliot_function}), $\mu$ is very large and we get that
\begin{equation}
 C_\phi =\frac{\mu}{U} \rightarrow  \frac{1 + n}{3+n}.
\end{equation} 
During this time $\dot{\phib}$ is very subdominant and we may assume that $\frac{\dot{b}}{b} \approx \frac{\dot{a}}{a}$. The Friedmann equation then gives
\begin{equation}
 3K_F \frac{\dot{a}^2}{a^2} =  8 \pi G a^2 e^{-4\phib_i} \bar{\rho}_r
\end{equation}
where $\phib_i$ is the initial condition of $\phi$ and $\bar{\rho}_r$ is the radiation energy density. Define
\begin{equation}
\Omega_{0r} = \frac{8\pi G \rho_{0r}e^{-4\phib_i}}{3 K_FH_0^2}
\end{equation}
where $\rho_{0r}$ is the proper radiation density today (as given by the radiation temperature), $H_0$ is the Hubble constant today, and 
the solution to the Friedman equation during radiation era is
\begin{equation}
 a =  \sqrt{\Omega_{0r}}H_0 t
\end{equation}

Although $\dot{\phib}$ is subdominant, we expect it to grow in order to approach the tracker solution. We therefore assume the ansatz
\begin{equation}
  \phi = \phi_i - \phi_1 a^m
\end{equation}
which gives
\begin{equation}
\dot{\phib} = -m\phi_1 a^m \frac{\dot{a}}{a}
\end{equation}
where $\phi_1$ and $m$ are constants. I assume that initially $\dot{\phib} = 0$.
Now  the variable $q = -2\mu \dot{\phib}$ evolves as
\begin{equation}
 \frac{dq}{d\ln a} + 3 q = 6 K_F \frac{\dot{a}}{a}
\end{equation}
which gives the evolution of $\mu$ as
\begin{equation}
 \frac{d\mu}{d\ln a} + (m+1)\mu = \frac{3}{m\phi_1} K_F e^{-m\ln a}
\end{equation}
The solution is
\begin{equation}
 \mu = \frac{3}{m\phi_1} K_F  a^{-m}
\end{equation}
so $\phi_1>0$
and we get
\begin{equation}
  \mu = \frac{3 K_F }{m(\phi_i - \phi)}
\end{equation}
irrespective of $\ell_B$ and $\mu_0$.
We now use the TeVeS constraint on $\dot{\phib}$ for the large $\mu$  to get
\begin{equation}
  \dot{\phib}^2 = \frac{\mu_0^2}{32\pi \ell_B^2} a^2 e^{-2\phib_i} \hat{\mu}^{1+n}
\end{equation}
Using the solution $a(t)$ to get $\phi(t)$ and $\mu(t)$ and then use the constraint we get
\begin{equation}
 m = \frac{4}{3+n}
\end{equation}
and
\begin{equation}
\phi_1   =\frac{1}{m}\left[ \frac{ e^{-2\phib_i}  (3K_F)^{1+n} }{32\pi \ell_B^2  \mu_0^{n-1} \Omega_{0r} H_0^2   } \right]^{1/(3+n)}
\end{equation}
For the TeVeS case, i.e. $n=2$, and $\phib_i = 0$,
  we get the Dodelson-Liguori~\cite{DL} result $m=4/5$ and $\phi_1 = \frac{5}{4}  \left[ \frac{27}{32\pi \ell_B^2 \Omega_{0r} H_0^2 \mu_0}\right]^{1/5}$.

\section{Cosmological perturbations}
\label{PT}
The perturbed equations are written directly in Fourier space, and are thus dependent on the wave number $k$.

Let us first define the following constants : 
\begin{eqnarray}
 K_t &=& K_B + K_+ - K_A \\
 \kappa_d &=& K_+ + \frac{1}{2} K_0 \\
 K_F &=& 1 + K_0 + \kappa_d \\
 R_K &=& 1 - \frac{3\kappa_d}{K_F}
\end{eqnarray}
As it turns out, for scalar and tensor modes $K_A$ never appears by itself to linear order and is always absorbed in to the combination $K_t$ defined above.
Moreover the constant $\kappa_d$ functions as a damping constant in the scalar mode vector field equation when expressed in gauge invariant variables. When $\kappa_d = 0$,
the vector field equation is independent of $k$. Thus $\kappa_d $ must be positive and very close to zero, for if it were negative, the perturbations would have
negative square speed of sound on flat space and thus be greatly unstable on small scales, 
while if it were large and positive, it would damp the vector field on cosmological scales, which
would render the theory irrelevant for structure formation. The constants $K_t$ and $R_K$ must be obey $0<K_t<2$ and $0<R_K\le 1$  respectively
for the energy densities  and the square speed of sound of all modes  to be positive (the last conditions are sufficient to ensure that $\kappa_d\ge0$).
Thus scalar modes depend on $K_t$, $K_F$ and $R_K$ and tensor modes on $K_F$ and $R_K$ only.

For vector modes the situation is slightly different and all constants $K_i$ are needed. In that case one can use the parameters $K_t$, $K_F$ and $R_K$ defined above
but one more is needed which can be either $K_A$ or $K_B$ which must obey $K_B \ge \frac{K_F R_K - 1}{2K_F R_K }$.

\subsection{Scalar modes}
Scalar modes are defined as in~\cite{skordis_PT}. 
The scalar field is perturbed as $\phi = \phib + \varphi$. The vector field has only one scalar mode, $\alpha$, because the other one is fixed by the timelike constraint
  and is defined as $A_i = a e^{-\phib} \grad_i \alpha$.
The matter frame metric has four scalar modes $\Xi$, $\chi$, $\zeta$ and $\nu$ such that $g_{00} = -a^2(1 - 2 \Xi)$, $g_{0i} = -a^2 \grad_i \zeta $, $g_{ij} = a^2 (1 + \frac{1}{3} \chi) q_{ij} + a^2( q_i^{\;\;k} q_j^{\;\;l} - \frac{1}{3} q_{ij} q^{kl})\grad_k \grad_l \nu $.  The
 Einstein frame metric has also four scalar modes $\XiE$, $\chiE$, $\zetaE$ and $\nuE$ defined in a similar way, related to the matter frame metric as
$\XiE = \Xi + \varphi$, $\chiE = \chi+ 6\varphi$, $\zetaE = \zeta - (1 - e^{-4\phib})\alpha$  and $\nuE = \nu$. 
The fluid variables are  the density contrast $\delta = \frac{\delta \rho}{\bar{\rho}}$
and momentum divergence $\theta$ such that the fluid velocity perturbation is defined as $u_i = a  \grad_i \theta$, where $u_a$ is the unit-timelike wrt $g_{ab}$ fluid velocity.

\subsubsection{Fluid equations}
The density contrast evolves as
\begin{equation}
\dot{\delta} = 3\frac{\dot{a}}{a}\left( w - C_s^2\right) \delta + \left(1 + w \right) \left(-k^2 \theta - \frac{1}{2} \dot{\chi} + k^2 \zeta\right)
\label{GFI_delta}
\end{equation}
where  $C_s^2 = \frac{\delta P }{\delta \rho}$ is the speed of sound.

The momentum divergence evolves as
\begin{eqnarray}
\dot{\theta} = -\Xi  + \frac{\dot{a}}{a} (3w - 1 )\theta  + \frac{C_s^2}{1+w} \delta  - \frac{\dot{w}}{1+w}\theta - \frac{2}{3}\left(k^2  - \curv{3}\right)\Sigma
\label{GFI_theta}
\end{eqnarray}
where $\Sigma$ is the fluid's scalar anisotropic stress (see ~\cite{skordis_PT}).

\subsubsection{Scalar field equations}
The scalar field perturbation evolves as
\begin{equation}
 \dot{\varphi} = - \frac{C_\phi}{2\mu\dot{\phib}} \gamma - \dot{\phib} \XiE
\label{GFI_varphi}
\end{equation}
where the auxiliary scalar field perturbation $\gamma$~\footnote{This differs from $\gamma$ variable in ~\cite{skordis_PT} by a multiplication with $a e^{-\phib} \dot{\phib}$ } evolves as
\begin{eqnarray}
\dot{\gamma}&=& 
 - \left[ (1 + 3C_\phi)\frac{\dot{b}}{b} + 4\dot{\phib}  +  8\pi G a^2 e^{-4\phib} \frac{C_\phi}{2\mub\dot{\phib}} (\rhob + 3 \Pb)  \right] \gamma
 + \mub \dot{\phib} k^2 e^{-4\phib} \left[ \varphi + \dot{\phib} \alpha\right]
\nonumber
\\
&&
 + \mub\dot{\phib}^2 \left[\dot{\chiE} - 2k^2 \zetaE \right]
+ 8\pi G a^2 e^{-4\phib} \rhob \dot{\phib} \left[ (1 + 3 C_s^2)\delta - (1+3w)(\XiE + 2 \varphi) \right]
\label{GFI_gamma}
\end{eqnarray}

\subsubsection{Vector equation}
The vector field equation is
\begin{equation}
 \dot{\alpha}  = E - \XiE + \left(\dot{\phib} -  \frac{\dot{a}}{a} \right)\alpha
\label{GFI_alpha}
\end{equation}
where the auxiliary gauge invariant vector mode $E$ evolves as
\begin{eqnarray}
 &&  K_t \left[ \dot{E} + \frac{\dot{b}}{b} E \right]
+    \curv{} \left(1 - K_F e^{4\phib}\right) \left(  \dot{\nu} + 2 (\zeta - \alpha) \right)
-    e^{4\phib}  \kappa_d \left(k^2 - \curv{3}\right)\left[\dot{\nu} +  2\left(\zeta - \alpha\right)    \right]
\nonumber
\\
&&
+   \frac{1}{3}(K_F e^{4\phib}- 1)  \left[  \dot{\chiE} + k^2 \dot{\nu}+ 6 \frac{\dot{b}}{b}  \XiE 
+  6\left(-\frac{\ddot{b}}{b} +2\frac{\dot{b}^2}{b^2} -2 \frac{\dot{b}}{b} \dot{\phib}\right) \alpha
\right]
+   ( 2e^{4\phib}-1) \bar{\mu}\dot{\phib}\left(\varphi - \dot{\phib}\alpha\right)
= 0
\label{GFI_E}
\end{eqnarray}
The coupling to the matter velocity in the equation above has been eliminated with the use of (\ref{GFI_ETheta}) below.

\subsubsection{Einstein equations}
The two Einstein constraint equations are
\begin{eqnarray}
&&
  \frac{\dot{b}}{b} K_F\left[ \dot{\chiE} + 2k^2(\alpha - \zeta) + 6\frac{\dot{b}}{b}\XiE\right]
 + \frac{1}{3}\left(k^2 - \curv{3}\right)e^{-4\phi}\left[ \chiE + k^2\nu\right]
\nonumber
=
\\
&&
 8\pi G a^2 e^{-4\phi} \bar{\rho} \left[\delta - 2\varphi \right]
 +e^{-4\phi}k^2 (K_t  E + 2\frac{\dot{b}}{b}\alpha)
 - \gamma 
\label{GFI_EDelta}
\end{eqnarray}
and
\begin{eqnarray}
 - K_F\left[  \frac{1}{3}  \left(\dot{\chiE} + k^2 \dot{\nu}\right) + 2 \frac{\dot{b}}{b}  \XiE \right]
&=&
 8\pi G a^2 e^{-4\phi}\left(\bar{\rho} + \bar{P}\right)\theta  
+ 2\bar{\mu}\dot{\phib}\varphi
 -\curv{2} e^{-4\phib} \alpha 
\nonumber 
\\
&&
-  \left[ K_F \curv{} + \kappa_d \left(k^2 - \curv{3}\right) \right]\left[\dot{\nu} + 2(\zeta -\alpha)  \right]
\label{GFI_ETheta}
\end{eqnarray}
while the propagation equations are
\begin{eqnarray}
&&
- K_F\bigg\{\ddot{\chiE} + 2k^2(\dot{\alpha}  - \dot{\zeta})
+ 6\frac{\dot{b}}{b}\dot{\XiE}+  6 \left[ 2\frac{\ddot{b}}{b} - \frac{\dot{b}^2}{b^2} + 4\dot{\phib}\frac{\dot{b}}{b}\right]\XiE  
+ 2(\frac{\dot{b}}{b} + \dot{\phib})\left[\dot{\chiE} + 2k^2(\alpha-\zeta) \right]
\bigg\}
\nonumber
\\
&&
- \frac{1}{3}e^{-4\phib}(k^2 - \curv{3})(\chiE + k^2\nu)
+ 3C_\phi \gamma
=
24\pi G a^2 e^{-4\phib} \bar{\rho} ( C_s^2 \delta - 2w\varphi)
- 2k^2 e^{-4\phib}E - 2k^2 e^{-4\phib} \frac{\dot{b}}{b}\alpha
\label{GFI_EP}
\end{eqnarray}
for the coupling to the perturbed pressure and
\begin{eqnarray}
 K_F R_K \left[    \ddot{\nu}
+ 2 ( \dot{\zeta} -  \dot{\alpha})
 + 2\left( \dot{\phib} + \frac{\dot{b}}{b}\right)\left(\dot{\nu} + 2 \zeta - 2\alpha\right)
\right]
+ 2 e^{-4\phib} E
+  e^{-4\phib} \left[ 2\frac{\dot{b}}{b}\alpha
- \frac{1}{3}  ( \chiE + k^2\nu)
\right]
 &=& 
16\pi G a^2 e^{-4\phib} (\bar{\rho} + \bar{P})\Sigma
\label{GFI_ESigma}
\end{eqnarray}
for the coupling to the shear.

\subsection{Vector modes}

Vector modes are defined as in~\cite{skordis_PT}. All vector modes have two polarizations and are purely spacial and divergenless.
 The vector field has a vector mode $\beta_i$ and is defined as $A_i = a e^{-\phib} \beta_i $.
The matter frame metric has two vector modes $r_i$, and $f_i$ such that $g_{0i} = -a^2 r_i $ and $g_{ij} = 2a^2\grad_{(i} f_{j)} $.  The
 Einstein frame metric has also two vector modes $\tilde{r}_i$, and $\tilde{f}_i$ defined in a similar way, related to the matter frame metric as
 $\tilde{r}_i = r_i - (1 - e^{-4\phib})\beta_i$  and $\tilde{f}_i = f_i$. 
The fluid variable is the vector mode $v_i$ in the fluid momentum such that the fluid velocity perturbation is defined as $u_i = a  v_i$.

\subsubsection{Fluid equations}
The fluid vector mode $v$ evolves as
\begin{equation}
\dot{v} = -\left[ (1 - 3w)\frac{\dot{a}}{a} + \frac{\dot{w}}{1+w}\right] v -\left(k^2 -\curv{2}\right) \sigma^{(v)}
\end{equation}
where $\sigma^{(v)}$ is the fluid's vector anisotropic stress (see ~\cite{skordis_PT}).

\subsubsection{Vector field equations}
The vector mode equation is
\begin{equation}
\dot{\beta} =  \epsilon + \left(\dot{\phib} - \frac{\dot{a}}{a}\right) \beta
\end{equation}
while the auxiliary vector mode $\epsilon$ evolves as
\begin{eqnarray*}
&& - K_t\left(\dot{\epsilon}  + \frac{\dot{b}}{b}\epsilon\right)
- \frac{1}{2} \left[ 1 + ( 2K_B - 1) e^{-4\phi}  \right]  \left(k^2  + \curv{2}\right)  \beta 
+  \frac{1}{2} \left( 1 -  K_F R_K e^{4\phib} \right) \left(k^2  - \curv{2}\right)   \left(\dot{f}+r -\beta \right) 
\\
&&
+   \left( K_F  e^{4\phi} - 1 \right) \left( 2 \frac{\ddot{b}}{b} - 4 \frac{\dot{b}^2}{b^2} + 4 \frac{\dot{b}}{b} \dot{\phib} \right) \beta 
+   \left( 2e^{4\phi} - 1\right) \mu \dot{\phib}^2 \beta
=0
\end{eqnarray*}
Notice that unlike scalar modes, the parameter $K_A$ is no longer redundant. In this case we can parametrize the vector field with $K_t$, $K_B$, $K_F$ and $R_K$.

\subsubsection{Einstein field equations}
We have the constraint equations
\begin{eqnarray*}
  \left(k^2  - \curv{2}\right)\left[  K_F R_K \left(\dot{f}+r -\beta \right) +    e^{-4\phib}\beta 
\right]
& = & 
  - 16\pi G a^2 e^{-4\phib} (\rhob + \Pb) v
\end{eqnarray*}
and the propagation equation
\begin{eqnarray*}
 K_F R_K \left[   \ddot{f} + \dot{r} -  \dot{\beta}
+ 2  \left(\frac{\dot{b}}{b} + \dot{\phib}\right) \left(  \dot{f}  +  r  - \beta \right)
\right]
 + e^{-4\phib}  \left(\dot{\beta} + 2 \frac{\dot{a}}{a}\beta\right)
 = 16\pi G a^2 e^{-4\phib}  (\rhob + \Pb) \sigma^{(v)}
\end{eqnarray*}

\subsection{Tensor modes}
Tensor modes are defined as in~\cite{skordis_PT}. All tensor modes have two polarizations and are purely spacial and divergenless.
The matter frame metric has a tensor mode $H_{ij}$,such that  $g_{ij} = a^2 H_{ij} $.  The
 Einstein frame metric has also a tensor mode $\tilde{H}_{ij}$ defined in a similar way, related to the matter frame metric as
$\tilde{H}_{ij} = H_{ij}$. 
The equation of motion for the tensor mode $H$ is
\begin{eqnarray*}
  K_F R_K \left[ \ddot{H} + 2 \left(\frac{\dot{b}}{b} + \dot{\phib}\right) \dot{H} \right]
  + e^{-4\phib} \left( k^2 + \curv{2}\right) H
=  16\pi G a^2 e^{-4\phib} (\rhob + \Pb)\sigma^{(T)}
\end{eqnarray*}
where $\sigma^{(T)}$ is the fluid's tensor anisotropic stress (see ~\cite{skordis_PT}).

\subsection{Adiabatic initial conditions for scalar modes}

\subsubsection{Conformal synchronous gauge in radiation era}
We start by adopting the scalar mode perturbation equations, to the synchronous gauge (defined as $\Xi = \zeta = 0$, $\chi = h$ and $-k^2\nu = h + 6 \eta$), in the radiation era, using the
background solution discussed above. Lets also define the following dimensionless variables : $x = k t$, $v = k\theta$, $u = k\alpha$, $\sigma = \frac{2}{3}k^2\Sigma$
 and $y = \frac{\gamma}{k^2}$. All equations are then written in dimensionless form, where derivatives wrt $x$ are denoted by a prime. For simplicity
let us also define $\phi_r = m\phi_1 \Omega_{0r}^{m/2} H_k^m $ where $H_k = H_0/k$, such that $\phib' = -\phi_r x^{m-1}$ (see section-\ref{back_rad}).

We have the fluid equations for photons given by
\begin{equation}
\delta_\gamma' =  -\frac{4}{3}v_\gamma - \frac{2}{3} h'
\end{equation}
and
\begin{eqnarray}
v_\gamma' =   \frac{1}{4} \delta_\gamma 
\end{eqnarray}
where the photon shear as well as higher moments of the Boltzmann hierarchy are vanishingly small due to the tight-coupling of photons to baryons and are ignored.
Likewise the fluid equations for neutrinos are
\begin{equation}
\delta_\nu' =  -\frac{4}{3} v_\nu - \frac{2}{3} h'
\end{equation}
\begin{eqnarray}
v_\nu' =   \frac{1}{4} \delta_\nu - \sigma_\nu
\end{eqnarray}
and
\begin{eqnarray}
\sigma_\nu' =   \frac{4}{15}v_\nu + \frac{2}{15}(h' + 6\eta')
\end{eqnarray}
 where higher moments of the Boltzmann hierarchy are small because they are of higher powers in expansions about $x$ and are ignored.

The scalar field evolves according to
\begin{equation}
 \varphi' =  \frac{C_\phi}{6K_F} x y + \phi_r x^{m-1} \varphi
\label{varphi_p}
\end{equation}
and
\begin{eqnarray}
y' &=& 
 -  \frac{(1+2C_\phi)}{x} y
 - \frac{3K_F}{x}   e^{-4\phib_i} \varphi 
+\frac{3K_F}{x}\phi_r x^{m-1} \left[h' + 6\varphi'  \right]
+\frac{3K_F}{x}\phi_r x^{m-1} \left[ 2 -  e^{-4\phib_i} \right] u
\nonumber
\\
&&
- 6K_F \phi_r x^{m-3} \left[ \Omega_{0\gamma}\delta_\gamma + \Omega_{0\nu}\delta_\nu - 3 \varphi \right]
\label{gamma_p}
\end{eqnarray}

The vector field obeys
\begin{equation}
 u'  = E - \varphi -  \frac{1}{x} u
\end{equation}
and
\begin{eqnarray*}
 &&  K_t \left[ E' + \frac{1}{x} E \right]
+    e^{4\phib_i}  \kappa_d \left[h' +  2u  \right]
+   2(1- K_F R_K e^{4\phib_i})  \eta'
\\
&&
+   2(K_F e^{4\phib_i}- 1)  \left[  \varphi' +  \frac{1}{x}  \varphi +  \frac{2}{x^2}  u \right]
- \frac{3K_F}{x}  ( 2e^{4\phib_i}-1) \left(\varphi + \phi_r x^{m-1}  u \right)
= 0
\end{eqnarray*}
Finally we need the two Einstein constraint equations
\begin{eqnarray*}
  \frac{1}{x} K_F\left[ h' + 6\varphi'+ 2u + \frac{12}{x}\varphi\right]
&=&
 \frac{3K_F}{x^2} \left[\Omega_{0\gamma}\delta_\gamma +  \Omega_{0\nu}\delta_\nu \right]
 + 2 e^{-4\phib_i}\left[\eta -  \varphi \right]
 +e^{-4\phib_i} (K_t  E + \frac{2}{x}u ) - \gamma 
\end{eqnarray*}
and
\begin{eqnarray*}
 R_K\eta' 
&=&
 \frac{2}{x^2}(\Omega_{0\gamma} v_\gamma + \Omega_{0\nu} v_\nu)
+  \varphi'  -  \frac{2}{x}  \varphi 
+   \frac{\kappa_d}{2K_F}  \left[h' + 2u  \right]
\end{eqnarray*}

\subsubsection{Adiabatic ansatz}
The adiabatic mode is such that $\eta \rightarrow 1 $ for $x\rightarrow 0$ while all other perturbations vanish in this limit (regularity assumption).
The adiabatic mode ansatz  $\eta = 1 + \eta_2 x^2$, $h = h_2 x^2$  solves the matter equations to give $\delta_\nu = \delta_\gamma = - \frac{2}{3}h_2 x^2$,
$v_\gamma = - \frac{1}{18}h_2x^3$, $v_\nu =  - \frac{1}{18}h_2x^3 -  \frac{2}{45}(h_2 + 6\eta_2) x^3$ and $\sigma_\nu = \frac{2}{15}(h_2 + 6\eta_2)x^2$.
We seek solutions to the scalar and vector field variables which are regular as $x \rightarrow 0$.
All the scalar and vector field terms in the Einstein constraint equations are then subdominant to lowest order in $x$
and we can solve them to get $h_2 = \frac{e^{-4\phib_i}}{2K_F} $ and $\eta_2 =  \frac{10-15R_K - 4S_\nu }{6(15R_K +  4 S_\nu )} h_2$, 
where $S_\nu = \frac{\Omega_{0\nu}}{\Omega_{0\nu} + \Omega_{0\gamma}}$.

Now consider the scalar field equation (\ref{varphi_p}) where the first term  clearly dominates at early times because of the regularity
condition on $\varphi$ and $y$. Let $y = y_0 x^p$ for some power $p\ge0$ to leading order.
We can then solve  (\ref{varphi_p}) to get
\begin{equation}
  \varphi =  \varphi_0 x^{2+p} = \frac{C_\phi}{6(2+p)K_F} y_0 x^{2+p}
\end{equation}
and using the above solution along with the one already found for $h$ and $\delta$ into (\ref{gamma_p}) we get
\begin{eqnarray}
(1 + p + 2C_\phi)y_0 x^p &=& 
10K_F\phi_r h_2 x^m
 - 3K_F e^{-4\phib_i} \varphi_0 x^{2+p}
+18(3+p)K_F\phi_r \varphi_0 x^{m+p}  
\nonumber
\\
&&
+ 3K_F\phi_r\left[ 2 -  e^{-4\phib_i} \right] u x^{m-1}
\end{eqnarray}
Since all terms above save the last one are regular as $x\rightarrow 0$, then the $u$-term must also be regular which means that $u = u_0 x^l$ 
with $l+m>1$.
Since  $0<m<1$ and $p>0$ , we have that $2+p>m$ and $m+p>m$ which means that as $x\rightarrow 0$, the first term would always dominate over the 
second and third term and the above equation is reduced to
\begin{eqnarray}
(1 + p + 2C_\phi)y_0 x^p &=& 
10K_F\phi_r h_2 x^m
+3K_F\phi_r\left[ 2 -  e^{-4\phib_i} \right] u_0 x^{l+m-1}
\label{y_red_2}
\end{eqnarray}
Now, the vector field equations become
\begin{equation}
(l+1) u_0 x^{l-1}  = E_0 x^q - \varphi_0 x^{2+p} 
\label{alpha_red}
\end{equation}
and
\begin{eqnarray}
   K_t  (1+q) E_0 x^{q} 
+    2e^{4\phib_i}  \kappa_d h_2 x^2
+   4(1- K_F R_K e^{4\phib_i})  \eta_2 x^2
+   4(K_F e^{4\phib_i}- 1)  u_0 x^{l-1}
&=& 0
\label{E_red}
\end{eqnarray}
where I  have ignored the $\varphi$-terms because they are all $\sim x^{2+p}$ and so the $h$ and $\eta$-terms always dominate them as $x\rightarrow 0$.
I have also  kept only the leading $u$-term.
Using (\ref{alpha_red}) into (\ref{E_red}) we get
\begin{eqnarray}
\left[   K_t  (1+q)  +   \frac{4}{1+l}(K_F e^{4\phib_i}- 1)  \right] E_0 x^q 
+    2e^{4\phib_i}  \kappa_d h_2 x^2
+   4(1- K_F R_K e^{4\phib_i})  \eta_2 x^2
&=& 0
\end{eqnarray}
where once again the $\varphi$-term has been ignored as it is of higher order than the $h$ and $\eta$ terms.
Therefore consistency requires that $q=2$ from which we get that the $\varphi$-term is of higher order than the $E$-term in (\ref{alpha_red}) which gives $l=3$.
Hence, the $u$-term in (\ref{y_red_2}) is of higher order and we get $p=m$.

 Reconstructing the full solution by using the above powers and matching coefficients then gives
\begin{eqnarray*}
 h &=& \frac{e^{-4\phib_i}}{2K_F} x^2
\\
 \eta &=& 1 +  \frac{10 -15R_K- 4S_\nu}{6(15R_K  + 4S_\nu )}h
\\
\frac{4}{3}\delta_b =\delta_\gamma = \delta_\nu &=& -\frac{2}{3} h
\\
\theta_\gamma &=& -\frac{1}{18} ht 
\\
 \theta_\nu &=&   -  \frac{15 R_K + 8 + 4S_\nu}{18(15R_K  + 4S_\nu)}  ht
\\
 \sigma_\nu &=&   \frac{4}{3(15R_K +  4 S_\nu)} h
\\
 E &=& \frac{ 20(K_FR_K e^{4\phib_i} - 1) + 2(15R_K  + 4S_\nu)( 1 - e^{4\phib_i} K_F)     }{3(15R_K  + 4S_\nu )( 3 K_t   +   K_F e^{4\phib_i}- 1 )} h
\\
 \alpha &=& \frac{1}{4}E t
\\
\gamma  &=& \frac{5  m\phi_1 \Omega_{0r}^{m/2} H_k^m  e^{-4\phib_i} }{1 + m + 2C_\phi} k^2 x^m
\\
\varphi &=& \frac{C_\phi}{6(2+m)K_F}\gamma t^2
\end{eqnarray*}
Note that for standard TeVeS with $\phib_i=0$ we get that $E=0$ and $\alpha=0$ to this order. In this very special case $E = O(3)$ and $\alpha = O(4)$ and
depend on higher powers of $h$ and $\eta$.

\section{Conclusion}
I have formulated the cosmological equations both at the background and linear perturbation level for a version of TeVeS theory with a generalized vector
field action. Using an analytical solution to the background equations for a general family of scalar field functions I constructed the primordial adiabatic
perturbation. The most general type of regular primordial perturnation is studied elsewhere~\cite{iso}. These equations can be used to study large scale structure
for these theories, to check whether there are stable versions of TeVeS which can agree with observations.

\section{Acknoledgments}
Research at Perimeter Institute for Theoretical Physics is supported in part by the Goverment of Canada through
NSERC and by the Province of Ontario through MRI.


\begin{thebibliography}{99}
\bibitem{wmap} D. N. Spergel \etal, \apjs{\bf 170}, 377 (2007).

\bibitem{LSS} M. Tegmark \etal, \prd{\bf 74}, 123507 (2006)

\bibitem{super} P. Astier \etal, \apj{\bf 447}, 31 (2006).

\bibitem{particle_DM1}  G. Jungman, M. Kamionkowski and K. Griest, Phys.Rept. {\bf 267}, 195 (1996). 

\bibitem{particle_DM2}  G. Bertone, D. Hooper and J. Silk,  Phys.Rept.{\bf 405}, 279 (2005).

\bibitem{void} P. J. E. Peebles, \apj{\bf 557}, 495 (2001). 

\bibitem{abel}  Mahdavi \etal, \apj{\bf 668}, 806 (2007).

\bibitem{teves} J.~D.~Bekenstein, \prd {\bf 70}, 083509  (2004); J.~D.~Bekenstein, JHEP PoS (jhw2004) 012 (astro-ph/0412652).
\bibitem{sanders_stratified} R.~H.~Sanders,  {\it Astrophys. J.} {\bf 480}, 492 (1997); ibid.
\mnras {\bf  363}, 459 (2005).

\bibitem{aqual}  J.~D.~Bekenstein and M.~Milgrom, \apj {\bf 286}, 7 (1984).

\bibitem{MOND} M.~Milgrom, \apj {\bf 270}, 365 (1983); ibid. \apj {\bf 270}, 371 (1983);
	 ibid. \apj {\bf 270}, 384 (1983).

\bibitem{MOND_galaxy} R. H. Sanders and S. S. McGaugh, Ann.Rev.Astron.Astrophys.{\bf 40}, 263 (2002). 


\bibitem{SMFB} C.~Skordis, D.~F. Mota, P.~G. Ferreira \& C. Boehm, \prl {\bf 96}, 011301 (2006).

\bibitem{DL} S.~Dodelson and M.~Liguori, astro-ph/0608602.

\bibitem{mismatch1} A. Lue, R. Scoccimarro, and G. D. Starkman, \prd{\bf 69}, 044005 (2004).

\bibitem{mismatch2} E. Bertschinger, \apj{\bf 648}, 796 (2006).

\bibitem{zhang} P. Zhang \etal, \prl{\bf 99}, 141302 (2007).

\bibitem{schmidt} F. Schmidt, M. Liguori and S. Dodelson, \prd{\bf 76}, 083518 (2007).

\bibitem{lensing1} H-S Zhao \etal, \mnras {\bf 368}, 171 (2006)

\bibitem{lensing2} D-M Chen and H-S Zhao, \apjl{\bf 650}, L9 (2006).

\bibitem{kahya}  E. O. Kahya and R. P. Woodard, Phys.Lett.{\bf B652}, 213 (2007).

\bibitem{teves_single}T. G. Zlosnik, P. G. Ferreira and G. D. Starkman, \prd{\bf 74}, 044037 (2006).

\bibitem{ZFS1} T. G. Zlosnik, P. G. Ferreira and G. D. Starkman, \prd{\bf 75}, 044017 (2007).

\bibitem{einstein_ether} T.~Jacobson \& D.~Mattingly, \prd {\bf 64}, 024028 (2001); 
ibid. \prd {\bf 70}, 024003 (2004). 

\bibitem{ZFS2} T. G. Zlosnik, P. G. Ferreira and G. D. Starkman, ArXiv:0711.0520

\bibitem{carroll}  S. M. Carroll and E. A. Lim, \prd{\bf 70}, 123525 (2004).

\bibitem{lim} E. A. Lim, \prd{\bf 71}, 063504 (005).

\bibitem{LMB} B. Li, D. F. Mota and J. D. Barrow,  arXiv:0709.4581. 

\bibitem{bonvin} C. Bonvin \etal, arXiv:0707.3519

\bibitem{zhao} H-S. Zhao, \apjl{\bf 671}, L1 (2007).

\bibitem{halle_zhao} A. Halle and H-S. Zhao, arXiv:0711.0958. 

\bibitem{seifert} M. D. Seifert, \prd{\bf 76 }, 064002 (2007).

\bibitem{skordis_dom} C. Skordis, in preparation.

\bibitem{bekenstein_review} J. D. Bekenstein, Contemporary Physics {\bf 47}, 387 (2006).

\bibitem{mavromatos} N. Mavromatos and M. Sakellariadou, Phys.Lett. {\bf B652}, 97 (2007).

\bibitem{brunetton} J-P Brunetton and G. Esposito-Farese, \prd{\bf 76}, 124012 (2007).

\bibitem{skordis_PT} C.~Skordis,  \prd {\bf 96}, 011301 (2006).

\bibitem{iso} C.~Skordis, in preparation.

\bibitem{wald} R.~M~.Wald, {\it General Relativity}, The University of Chicago Press, Chicago and London, 1984. 

\bibitem{spher_sym_teves} D.~Giannios, \prd {\bf 71}, 103511 (2005). 

\bibitem{Bourliot} F.~Bourliot \etal, \prd {\bf 96}, 011301 (2006).

\bibitem{quinte}  B. Ratra $\&$ P. J. E. Peebles, \prd{\bf 37}, 3406 (1988);
C. Wetterich, Nucl. Phys. {\bf B252}, 302 (1988); ibid Nucl. Phys {\bf B252}, 668;
ibid. Astron. $\&$ Astroph. {\bf 301}, 321 (1995); E. Copeland \etal, Ann. N.Y. Acad. Sci. {\bf 688}, 647 (1993); P. G. Ferreira and M. Joyce, \prl{\bf 79}, 4740 (1997), 
 A. Albrecht and C.Skordis, \prl{\bf 84}, 2076 (2000); C.Skordis  and A. Albrecht, \prd{\bf 66}, 043523 (2002).

\end{thebibliography}
\end{document}